\documentclass[fleqn,usenatbib,usedcolumn]{mnras}

\usepackage[T1]{fontenc}
\usepackage{ae,aecompl}

\usepackage{graphicx}   
\usepackage{amsmath}    
\usepackage{amssymb}    
\usepackage{natbib}
\usepackage{enumerate}
\usepackage{txfonts}

\newcommand{\Msun}{$M_{\sun}$}

\def\aj{{AJ}}

\def\apj{{ApJ}}
\def\apjs{{ApJS}}
\def\mnras{{MNRAS}}

\def\pasp{{PASP}}
\def\s4g{{S$^4$G}}
\def\Msun{$M_\odot$}
\def\ser{S{\'e}rsic}
\def\mus{{{$\mu m$}}}

\def\mum{{{$\mu m$}}}

\usepackage{color}
\definecolor{lightred}     {rgb}{1.00, 0.35, 0.05}
\definecolor{darkred}      {rgb}{0.70, 0.05, 0.05}
\definecolor{lightblue}    {rgb}{0.05, 0.35, 1.00}
\definecolor{blue}         {rgb}{0.03, 0.25, 0.85}
\definecolor{darkblue}     {rgb}{0.05, 0.05, 0.70}
\definecolor{green}        {rgb}{0.05, 0.65, 0.15}
\definecolor{darkgreen}    {rgb}{0.10, 0.55, 0.15}

\title[Bar-Induced Secular Evolution in Galaxies]{Evidence of bar-induced secular evolution in the inner regions of stellar discs in galaxies: what shapes disc galaxies?}

\author[Kim et al.]{
Taehyun Kim,$^{1,2,3}$\thanks{E-mail: tkim@kasi.re.kr}
Dimitri A. Gadotti,$^{4}$
E. Athanassoula,$^{5}$
Albert Bosma,$^{5}$
\newauthor
Kartik Sheth,$^{2,6}$
Myung Gyoon Lee$^{3}$
\\
$^{1}$Korea Astronomy and Space Science Institute, Daejeon, Korea\\
$^{2}$National Radio Astronomy Observatory, 520 Edgemont Road, Charlottesville, VA 22903, USA\\
$^{3}$Astronomy Program, Department of Physics and Astronomy, Seoul National University, Seoul, Korea\\
$^{4}$European Southern Observatory, Casilla 19001, Santiago 19, Chile\\
$^{5}$Aix Marseille Universit{\'e}, CNRS, LAM (Laboratoire d'Astrophysique de Marseille) UMR 7326, 13388 Marseille, France\\
$^{6}$NASA Headquaters, 300 E. St SW, Washington DC 20546, USA
}

\date{Accepted XXX. Received YYY; in original form ZZZ}

\pubyear{2016}

\begin{document}
\label{firstpage}
\pagerange{\pageref{firstpage}--\pageref{lastpage}}
\maketitle

\begin{abstract}
We present evidence of bar-induced secular evolution in galactic discs using 3.6 {\mum} images of nearby galaxies from the {\it Spitzer} Survey of Stellar Structure in Galaxies ({\s4g}). 
We find that among massive galaxies ($M_{\ast}/${\Msun}$> 10^{10}$), longer bars tend to reside in inner discs having a flatter radial profile.
Such galaxies show a light deficit in the disc surrounding the bar, within the bar radius and often show a $\Theta$-shaped morphology.
We quantify this deficit and find that among all galaxies explored in this study (with $10^{9}<M_{\ast}/${\Msun}$< 10^{11}$), galaxies with a stronger bar (i.e. longer and/or with a higher Bar/T) show a more pronounced deficit.
We also examine simulation snapshots to confirm and extend results by Athanassoula and Misiriotis, showing that as bars evolve they become longer, while the light deficit in the disc becomes more pronounced.
Theoretical studies have predicted that, as a barred galaxy evolves, the bar captures disc stars in its immediate neighbourhood so as to make the bar longer, stronger and thinner. Hence, we claim that the light deficit in the inner disc is produced by bars, which thus take part in shaping the mass distribution of their host galaxies.
\end{abstract}




\section{Introduction}

The evolution of galaxies is initially driven by fast and violent processes at early times. Later on, however, as the merger rate drops, slow and secular processes start to become dominant (e.g., \citealt{kormendy_04}). 
One of the major contributors that drive internal secular evolution in disc galaxies are non-axisymmetric structures, such as bars (\citealt{athanassoula_13_book, kormendy_13_book, sellwood_14_rev} for reviews covering the theoretical and observational perspectives).
Bars are common in nearby disc galaxies (a fraction of 50$\sim$ 70 per cent, e.g., \citealt{buta_15} and references therein, a fraction of $\sim$30 per cent when only strong bars are counted. e.g., \citealt{masters_11, buta_10, buta_15} and references therein). 
Numerical and analytic studies show that once disc galaxies are massive enough and rotation-dominated, the bar instability develops relatively fast, within a few hundred Myr (e.g., \citealt{pfenniger_91, friedli_93, athanassoula_02a}, hereafter AM02; \citealt{martel_13}). 
However, the bar formation can be delayed if the disc is dispersion-dominated, if the initial dark matter halo is dominant, or if the galaxy initially contains a large amount of gas (\citealt{athanassoula_86}; AM02; \citealt{athanassoula_03, athanassoula_13}).
Thus the fraction of barred galaxies provides us statistics on how many galaxies already have sufficiently massive discs dominated by rotation (\citealt{sheth_12}). 
Bar fractions appear to change with redshift: from $z=0.8$ to $z=0.2$, overall bar fractions increase from 20 to 65 per cent, while strong bar fractions increase from 10 to 30 per cent (\citealt{sheth_08, cameron_10}, see also \citealt{melvin_14}). 
The lower bar fraction at high redshift is mainly due to the lower mass galaxies not yet having developed bars (\citealt{sheth_12}).
At z $>$1, there are tentative bar detections in the near infrared data (\citealt{sheth_03}), and the bar fractions appears to be $\sim 10$ per cent at z=1.5$\sim$2 (\citealt{simmons_14}).
Note that finding bars is subject to limits due to image resolution and band-shifting.

Cosmological simulations suggest that bars formed during the violent phase (z$>$1) may be easily destroyed or may become too weak to be observed, whereas bars formed in the stellar disc at the secular phase ($z<0.8$) are predicted to be generally robust and long-lived. The fraction of bars increases with time (\citealt{kraljic_12, martig_12}), consistent with the observed trend.

It is challenging to gauge when bars are formed, as the age of stellar populations building up a bar does not necessarily refer to the formation epoch of the bar itself. 
Nevertheless, several attempts have been made (e.g., \citealt{gadotti_05, wozniak_07, perez_07, perez_11, elmegreen_09, sanchez_blazquez_11, delorenzo_caceres_13, james_16}).  The age of bars seems to be correlated to the mass and dynamical status of their parent galaxies. Massive, and rotation dominated galaxies form their bars first (\citealt{sheth_08, sheth_12}).
Recent studies find that some galaxies have hosted their bars for a long time.
An integral field spectrograph study on NGC 4371 reveals that the inner disc and nuclear ring -- thought to be composed of stars formed from gas funnelled by the bar -- are mainly composed of old ($>10$ Gyr) stellar populations, and indicate that the bar was formed at $z\sim1.8$  (\citealt{gadotti_15}).

Although not a direct measure of bar age, radial light profiles of bars have been used to infer the dynamical age of bars (\citealt{kim_15a}).
Because bars are formed from disc material, the radial profiles of bars at an early evolutionary stage would be similar to those of discs, which are exponential (AM02). However, bars with exponential profiles are not necessarily young (AM02). 
Whereas, bars that show flat radial profiles are expected to be strong, and therefore dynamically old. 
Bars in massive and bulge-dominated galaxies are found to show flat radial profiles (\citealt{kim_15a}).
Numerical simulations find that once bars are formed, it is difficult to dissolve them (e.g., \citealt{shen_04, athanassoula_05b, athanassoula_13, debattista_06, berentzen_07, villa_vargas_10, kraljic_12, martig_12}).
Hence, bars must have been influencing their host galaxies since they are formed, and have an extended impact in the evolution of galaxies (\citealt{gadotti_15}).
Therefore, it is important to explore the impact of bar driven secular evolution on their host galaxies.

Because of their non-axisymmetry, bars induce large-scale streaming motions (e.g., \citealt{athanassoula_92a}). Observational studies find enhanced central gas concentrations in barred galaxies (\citealt{regan_97, sakamoto_99, sheth_00, zurita_04, jogee_05}). This leads to barred galaxies also showing enhanced star formation in the central region (e.g, \citealt{ho_97, ellison_11}), and pronounced nuclear rings (\citealt{knapen_02, comeron_10, kim_w_12a, seo_13}). Eventually, bars induce the formation of discy pseudo bulges in their host galaxies (\citealt{kormendy_04, sheth_05, athanassoula_05a, debattista_06, cheung_13}). 

The impact of bars on disc galaxies is not only limited to the central part of the galaxy. Bar torques bring the gas inside the corotation inwards, while pushing the gas between the corotation radius and the outer Lindblad resonance radius outwards (\citealt{combes_85, combes_08_gas, kubryk_13}).
Barred galaxies are often accompanied by an outer ring where one of the bar resonances is expected to be located (\citealt{schwarz_81, buta_96, buta_03, buta_15, romero_gomez_06, athanassoula_09a}). 
Thus bars drive secular evolution in their host galaxies, slowly re-arranging mass and angular momentum distributions throughout the different galactic components.

Numerical simulations predict that bar-induced angular momentum redistribution leads discs to show a break in their radial density profile (\citealt{valenzuela_03, debattista_06}) with a shallower inner disc and a steeply decreasing outer disc. Observational data also show that, compared to unbarred galaxies, barred ones show larger global disc scale length and fainter central surface brightness of discs among massive galaxies (\citealt{sanchez_janssen_13, diaz_garcia_16xx}). 
The majority of disc galaxies are found to show at least one disc break (e.g., \citealt{pohlen_02, pohlen_06, erwin_08, gutierrez_11, maltby_12a, munoz_mateos_13, kim_14, laine_14}; for edge-on galaxies, see also \citealt{comeron_12, martin_navarro_12}), and this has been confirmed by simulations (e.g., \citealt{roskar_08a, athanassoula_16b}). However, it should be noted that not all disc breaks are produced by bars, and even unbarred galaxies may show disc breaks. Apart from the bar-driven one, several other mechanisms responsible for disc breaks have been proposed (e.g., \citealt{vanderkruit_87, tagger_87, kennicutt_89, laurikainen_01, elmegreen_07, younger_07, roskar_08b, minchev_12a, munoz_mateos_13}).

If a galaxy has a disc break, there are two different disc scale lengths: the inner and the outer disc scale length. 
Compared to the global disc scale length which ignored the disc break, the inner and outer disc scale lengths differ by $\sim40$ per cent on average (\citealt{kim_14}). It is therefore important to separate a disc into its inner and outer parts. 
\cite{laine_14} find that the average disc profile of Type I is similar to the average outer disc profile of Type II, while inner disc profiles of Type II are flatter than disc profiles of Type I, and \citet{laine_14} attribute this difference to the effects of bars.

\begin{figure}
	\centering
	\includegraphics[width=7cm]{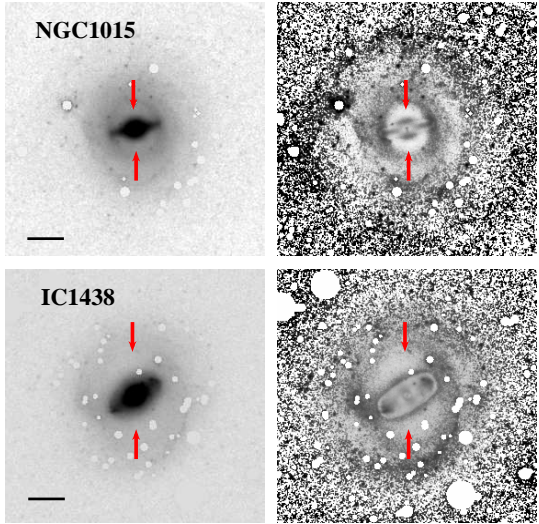}
	\caption{(Left): Images of NGC 1015 and IC 1438 at 3.6 {\mum}. Red arrows indicate the deficit of light from the inner disc surrounding the bar. Horizontal bars in the bottom left corners of these panels span 30 arcsec. North is up and east is to the left. 
(Right): Residual images using the models fitted in Paper I highlight the light deficits in the inner discs. 
}
	\label{fig:deficit_light}
\end{figure}

A morphology often seen in barred galaxies is well represented by the so-called $\Theta$-shaped galaxies, which show a light deficit in the disc surrounding the bar at radii smaller than the radius of the inner ring (i.e. the ring that is often present near the ends of the bar), and thus within the bar radius.
\citet{gadotti_03} have studied two of such barred galaxies (NGC 4608 and NGC 5701; see also \citealt{laurikainen_05, gadotti_08}). We show more examples in Fig.~\ref{fig:deficit_light}, in which arrows indicate the disc light deficit around the bar within the bar radius (note that the presence of an inner ring is not a necessary condition for the occurrence of the light deficit). 
Theoretical work also finds light deficits in simulated galaxies (e.g., AM02; \citealt{athanassoula_13}). In particular, a model with a more centrally concentrated halo shows a more prominent light deficit, as well as a stronger, longer and thinner bar, as compared to a model with a less centrally concentrated halo (AM02).

Theoretical studies have also predicted that bars give up angular momentum as they evolve, which leads to several changes in bar properties that might also affect the disc.
\citet{athanassoula_03}, considering the bar as an ensemble of orbits, presents schematically three possible changes in its orbital structure.
Firstly, the bar traps stars which were initially on quasi-circular orbits just outside the bar. The new orbits are elongated and thus the bar becomes longer and/or more massive, to the detriment of the surrounding disc. 
Secondly, orbits in the bar become more elongated, i.e. the bar becomes thinner. Lastly, the bar slows down by lowering its pattern speed. These three possible processes are closely linked together and occur simultaneously.
In summary, simulations predict that as bars evolve disc stars are captured onto bar orbits. With the help of these newly captured stars the bar becomes longer and more massive. 
Thus, since stars are removed from the disc and added to the bar, the inner part of the disc surrounding the bar (i.e. at galactocentric radii below the bar radius, or $r< R_{\rm bar}$) should in principle become less dense. Hence, we should expect that there will be a deficit of light from the disc surrounding the bar, and, as the galaxy evolves, this light deficit in the inner disc will become more pronounced. 
However, this effect has not yet been explored with observational datasets and no direct comparisons with simulations have been reported either.
In this study, we test this hypothesis by checking whether there is a relation between the bar and the {\it inner} disc properties through detailed structural analysis, quantifying the light deficit in both an observational dataset and a set of simulation snapshots.

The paper is organized as follows. In \S 2 we give a brief overview of our data and data analysis. Results on the impact of bar-driven secular evolution on discs are presented in \S 3. We explore how bar properties change with time using the snapshots of a simulation in \S 4.  We discuss our results in \S 5, and summarize and conclude in \S 6.

\section{Data and Data Analysis}
We made use of 3.6 {{\mum}} images drawn from the {\it Spitzer} Survey of Stellar Structure in Galaxies ({\s4g}, \citealt{sheth_10}, for other data products of the {\s4g}, please see \citealt{munoz_mateos_15, salo_15, querejeta_15}).  
Since less affected by dust contamination (e.g., \citealt{meidt_12a}), 3.6{{\mum}} images are optimal to investigate the stellar mass distribution in galaxies.
We use structural parameters presented in Kim et al. (2014, Paper I) derived via two-dimensional image decomposition. Galaxies have been fitted with a bulge, a bar, a nuclear point source (if any), and a disc including a disc break (if any) using the BUlge Disk Decomposition Analysis code ({\sc BUDDA v2.2}, \citealt{gadotti_08}, \citealt{desouza_04}).
We choose galaxies that show down-bending disc profiles (i.e. Type II) from the galaxy sample of  Paper I, resulting in 118 nearby galaxies in the mass range of $10^9 - 10^{11}$ {\Msun} and a wide range of Hubble types (SB0 -- SBdm). For details on the sample and two dimensional decomposition analysis we refer the reader to Paper I.

\begin{figure}
\centering
\includegraphics[width=8cm]{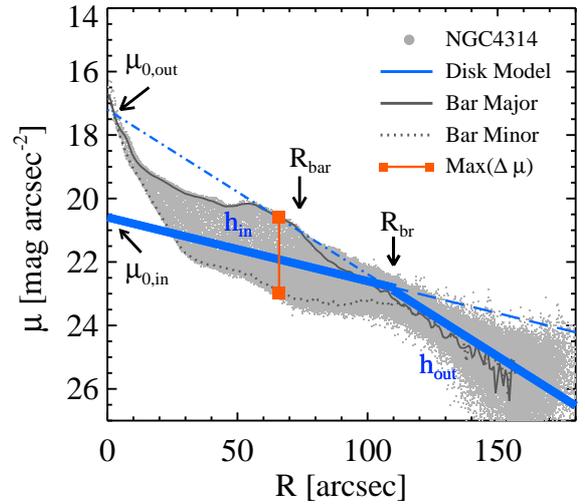}
\caption{
Surface brightness radial profile of NGC4314 at 3.6 {\mus}. A grey point represents a single pixel of the S$^4$G image. For simplicity, we only plot the model fit of the disc component, describing the disc break ($R_{\rm br}$), the scale length of the inner disc ($h_{\rm in}$), the scale length of the outer disc ($h_{\rm out}$), the central surface brightness of the inner disc component ($\mu_{\rm 0, in}$), and that of the outer disc component ($\mu_{\rm 0, out}$). The blue long-dashed line outlines the model fit of the inner part of the disc, and the blue dot-dashed line outlines that of the outer part of the disc, including their respective outward and inward extrapolations. 
Grey solid and dotted lines trace the surface brightness profile along the bar major and minor axis, respectively.
Orange vertical solid line denotes the maximum difference between the bar major and minor axis, Max($\Delta\mu$).
}
\label{fig:schematic}
\end{figure}

We separate the light emanating from the bulge, disc and bar components. Thus the central surface brightness of the disc is {\it not} the brightness at the centre of the galaxy, but that of the disc component from the model fit. 
In Fig.~\ref{fig:schematic}, we show as an example the fit of NGC 936. For simplicity, we only plot the model of the disc component even though we fit the galaxy also with a bulge and a bar component simultaneously. The disc of NGC 936 has a break at $R_{\rm br}\sim100$ arcsec, and the slope of the disc profiles inside and outside of the disc break are different, i.e. their disc scale lengths are different. We fit the light profile of the disc component using the following representation:

\begin{equation}
\begin{split}
\mu_{\rm disc}(r)= \begin{cases}
\mu_{\rm{0,in}} + 1.086 (r/h_{\rm in}),  ~ \mbox {if } r \leq R_{\rm br} 
\\
\mu_{\rm{0,out}} + 1.086 (r/h_{\rm out}), ~ \mbox {if } r > R_{\rm br}, 
\end{cases}
\label{eq:mu_wb}
\end{split}
\end{equation}

\noindent where $\mu_{\rm {0,in}}$ and $\mu_{\rm {0,out}}$ are the central surface brightness of the inner and the outer discs, and $h_{\rm {in}}$ and $h_{\rm out}$ are the scale length of the inner and the outer discs, respectively. 
Disc properties can thus be characterized by a disc scale length ($h$) and a central surface brightness ($\mu_{0}$), with which we can describe the slope of disc profile and the size of disc. 
If a disc has a large $h$ and low $\mu_0$, then the disc shows a flattened radial surface brightness profile. 
Conversely, the disc shows a steeply decreasing surface brightness profile if it has a small $h$ and high $\mu_0$.
$R_{\rm br}$ is the break radius and is fitted as a free parameter with BUDDA. 
In addition to the structural parameters described above, we also make use of the bar radius ($R_{\rm bar}$), the bar-to-total luminosity ratio (Bar/T), presented in Paper I, and the bar {\ser} index ($n_{\rm bar}$, from \citealt{kim_15a}). The stellar mass of the galaxy is estimated from the total magnitude at 3.6 {\mum}, and we also use $R_{\rm 25.5}$ (the radius where the surface brightness of the galaxy reaches 25.5 mag arcsec$^{-2}$ at 3.6 {\mum}); both are taken from \citet{munoz_mateos_15}.

\section{Observational Results}
\subsection{Properties of inner discs in barred galaxies}
\label{sec:res_surfb_hin}

\begin{figure*}
\centering
\includegraphics[width=16cm]{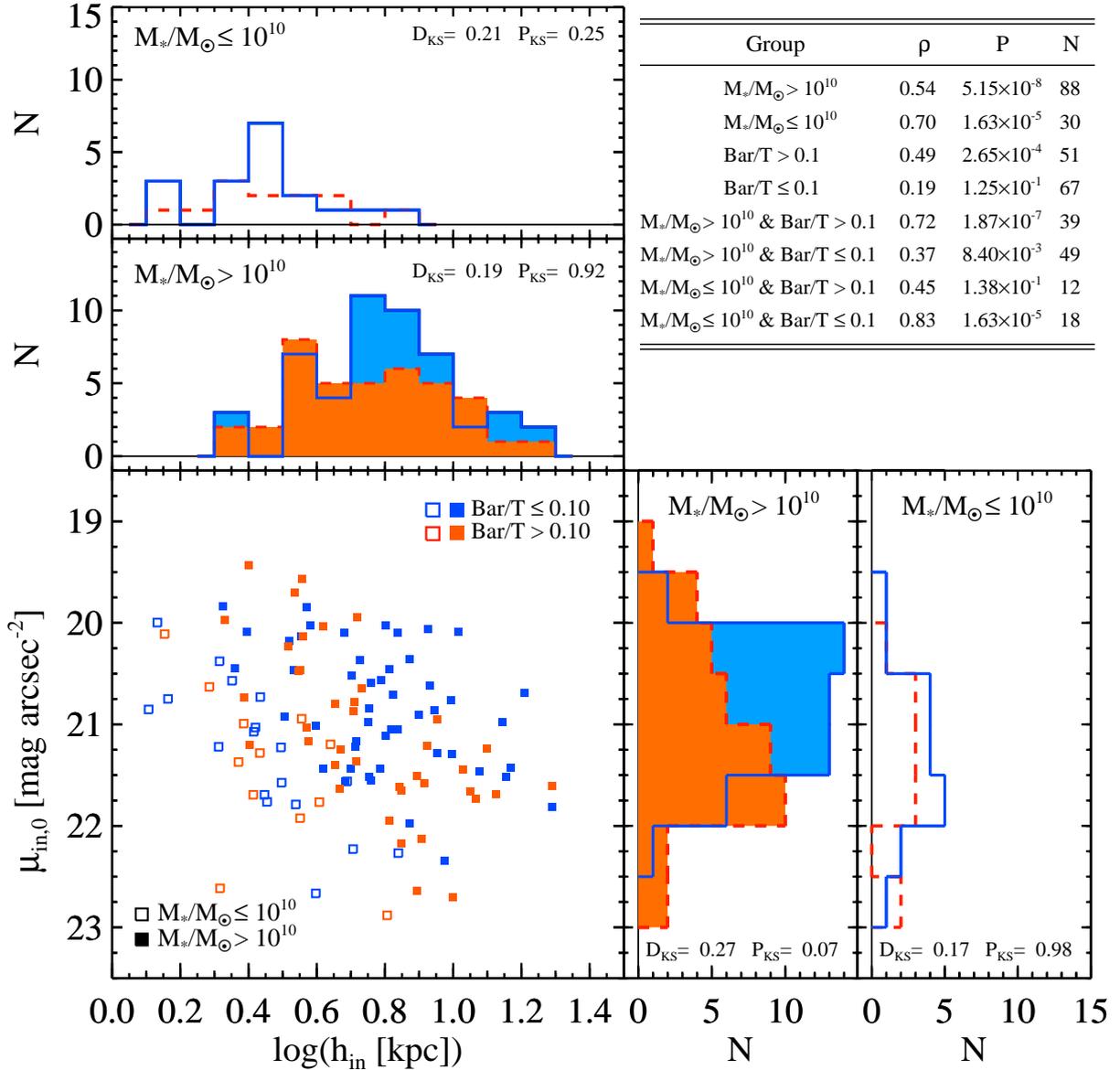}
\caption{Central surface brightness of inner discs ($\mu_{\rm 0,in}$) and inner disc scale length ($h_{\rm in}$) at 3.6 {\mus}. 
Note that the central surface brightness of the disc is {\it not} the brightness at the centre of the galaxy, but that of just the disc component.
Blue symbols denote galaxies with Bar/T $\leq$ 0.1 and red symbols denote galaxies with Bar/T $>$ 0.1. Massive galaxies ($M_{\ast}/${\Msun}$> 10^{10}$) are plotted with filled symbols, and less massive galaxies ($M_{\ast}/${\Msun}$\leq 10^{10}$) are in open symbols. 
The distributions of $\mu_{\rm 0,in}$ and $h_{\rm in}$ are also shown separately for galaxies with stellar masses above and below $10^{10}$ {\Msun}. 
In each panel of distributions, we present results from Kolmogorov--Smirnov test ($\rm P_{\rm KS}$ and $\rm D_{\rm KS}$).
In the upper right part, we give the Spearman correlation coefficients ($\rho$ and P) of the $\mu_{\rm 0}-{\rm log}(h_{\rm in})$ distributions and number of galaxies for each group.
}
\label{fig:mu0_hin}
\end{figure*}

We plot the central surface brightness of the inner disc ($\mu_{\rm 0,in}$) as a function of the inner disc scale length ($h_{\rm in}$) in Fig.~\ref{fig:mu0_hin}.
The disc central surface brightness is corrected for inclination using $\mu_{\rm cor}$ = $\mu_{\rm obs} - 2.5$log$(b/a)$, where $b/a$ is the axial ratio of the galaxy.
We split the sample into galaxies with Bar/T $>$ 0.1 and Bar/T $\leq$ 0.1, and plot them in red and blue, respectively. Massive galaxies ($M_{\ast} > 10^{10}${\Msun}) are shown in filled symbols, while less massive galaxies ($M_{\ast} \leq 10^{10}${\Msun}) are shown in open symbols. The Hubble types of both massive and less massive galaxies range from SB0 to SBdm (-2<T<10). The majority of massive galaxies are SBa -- SBbc (1$<$T$<$4), while the majority of less massive galaxies are SBc -- SBd (5$<$T$<$8).

Fig.~\ref{fig:mu0_hin} shows that the galaxies with a large $h_{\rm in}$ have a fainter $\mu_{\rm 0,in}$ on average. This is in agreement with previous studies on the disc scale relation that included disc break in the analysis (\citealt{laine_14}), and even with studies that did not include the disc break, but use a global disc scale length (e.g., \citealt{dejong_96III,  graham_01b, graham_01c, erwin_05_bar, gadotti_09, laurikainen_10, fathi_10b, sanchez_janssen_13}). 
Massive galaxies dominate the upper-right locus in the $\mu_{\rm 0,in}$--${\rm log}(h_{\rm in}/R_{\rm 25.5})$ plot in Fig.~\ref{fig:mu0_hin}. This is because massive galaxies show elevated surface brightness profiles, and have larger discs, on average. Therefore, they exhibit higher $\mu_{\rm 0,in}$ and larger $h_{\rm in}$ compared to less massive ones. 
There is a trend that early and intermediate type spirals are relatively more massive than late type spirals (e.g., \citealt{laurikainen_07}).
Therefore, this is in line with the studies that find late type spiral galaxies populate the lower left of the  $\mu_{\rm 0}$-- $h$ plane, while early and intermediate type galaxies spread over the diagram (e.g., \citealt{graham_01b, gadotti_09, fathi_10b}). 
In the upper right panel of Fig.~\ref{fig:mu0_hin}, we present the Spearman correlation coefficients ($\rho$) and the statistical significance of the correlation (P), number of samples for each group.

Our main interest in Fig.~\ref{fig:mu0_hin} is the distribution of $\mu_{\rm 0}$.
Interestingly, we find that massive galaxies with Bar/T $>$ 0.1 preferentially have lower $\mu_{\rm 0}$ than those with Bar/T $\leq$ 0.1. The distribution of $\mu_{\rm 0}$ for massive galaxies with Bar/T $>$ 0.1 is skewed and shows an increased number of galaxies that have a lower $\mu_{\rm 0,in}$.
We run the two-sided Kolmogorov--Smirnov (KS) test to examine whether the $h_{\rm in}$--, and $\mu_{\rm 0}$--distributions of the two groups (Bar/T $>$ 0.1 and Bar/T $\leq$ 0.1) are drawn from the same distribution. 
$\rm P_{\rm KS}$ is the probability that the two groups are drawn from the same parent distribution and $\rm D_{\rm KS}$ is the maximum deviation between the two cumulative distributions. $\rm P_{\rm KS}$ and $\rm D_{\rm KS}$ are presented in each panel. 
While $h_{\rm in}$ of the two groups (Bar/T  $>$ 0.1 and Bar/T  $\leq$ 0.1) are not significantly different, the KS-test shows that $\mu_{\rm 0}$ of the Bar/T $>$ 0.1 and Bar/T  $\leq$ 0.1 groups are clearly different among massive galaxies only, for which we got $\rm D_{\rm KS}$=0.27, $\rm P_{\rm KS}$=0.07 for $M_{\ast} >10^{10}${\Msun}.
This hints that bars might affect the inner part of the disc for massive and strongly barred galaxies (Bar/T $>$ 0.1). We will investigate this further in the next subsections. 
However, among less massive galaxies and galaxies with Bar/T $\leq$ 0.1, we find no clear trend. 


\begin{figure*}
\centering
\includegraphics[width=16cm]{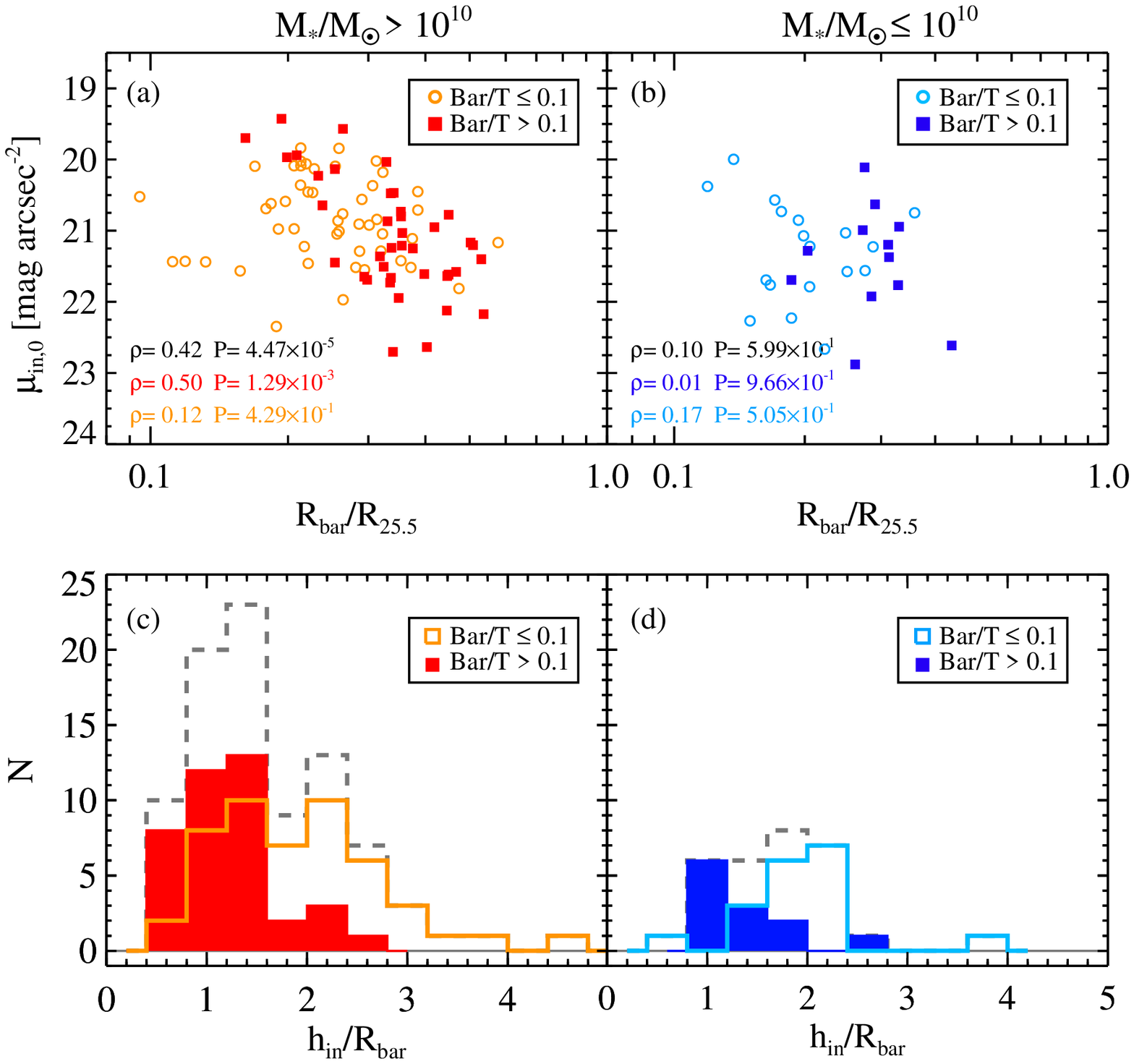}
\caption{
(a): Central surface brightness ($\mu_{\rm 0,in}$) of inner discs are plotted as a function of normalized bar radius $R_{\rm bar}/R_{\rm 25.5}$ for massive galaxies ( $M_{\ast} >10^{10}${\Msun}).
(b): The same as (a), but for less massive galaxies ($M_{\ast} \leq 10^{10}$ {\Msun}).
The galaxies with Bar/T $>$ 0.1 are plotted in filled squares, while the galaxies with Bar/T $\leq$ 0.1 are in open circles. $\rho$ and P are the Spearman correlation coefficient and the statistical significance of the coefficient, respectively.  
(c): Distribution of the ratio between inner disc scale length and bar radius ($h_{\rm in}/R_{\rm bar}$) for massive galaxies. The filled histogram represents galaxies with Bar/T $>$ 0.1 while the open histogram indicates galaxies with Bar/T $\leq$ 0.1. Dashed line shows the distribution of all massive galaxies. (d): The same as (c) but for less massive galaxies. 
}
\label{fig:mu0_hin2}
\end{figure*}

\subsection{Bar Length and Inner Disc Properties}
\label{sec:res2}
Next, we explore how the properties of the bar and the inner disc are linked to each other.
Specifically, we investigate how $h_{\rm in}$ and $\mu_{\rm 0}$ vary with the bar radius in Fig.~\ref{fig:mu0_hin2}. 
$\mu_{\rm 0}$ and the normalized bar radii ($R_{\rm bar}/R_{\rm 25.5}$) of massive galaxies are plotted in Fig.~\ref{fig:mu0_hin2}(a) and those of less massive galaxies are shown in Fig.~\ref{fig:mu0_hin2}(b).
$R_{\rm bar,0}$ is the deprojected bar radius, calculated analytically following the method developed in \citeauthor{gadotti_07} (2007, Appendix A) which assumes that the shape of the outermost part of the bar (thin part of the bar) can be well approximated by an ellipse.
Fig.~\ref{fig:mu0_hin2}(a) shows that among massive galaxies with Bar/T $>$ 0.1 (red filled squares), the galaxies with a longer bar tend to have fainter $\mu_{\rm 0}$.
We show the Spearman correlation coefficients $\rho$ and P in each panel.
Our results indicate a tight correlation ($\rho=0.5$) between $\mu_{\rm 0}$ and $R_{\rm bar}/R_{\rm 25.5}$, but for massive galaxies with Bar/T $>$ 0.1 only, at a significance level of 99.87 per cent, a result above $3\sigma$ significance.  
No such clear trend is found among galaxies with Bar/T $\leq$ 0.1, nor among less massive galaxies.
In Fig.~\ref{fig:mu0_hin2}(a), there are 6 galaxies offset to the lower left from the main trend. Those galaxies have very short bars compared to the size of the disc ($R_{25.5}$), and thus lie below the trend we found for the massive, higher Bar/T galaxies. These galaxies are found to be gas rich, and either bulgeless or have very insignificant bulges (Bulge/T$<$0.1). We will further investigate these in the upcoming paper. 
If we exclude these 6 galaxies, we obtain a mild correlation between $\mu_{\rm 0}$ and $R_{\rm bar}/R_{\rm 25.5}$ among massive and Bar/T $<$ 0.1 galaxies ($\rho=0.4, P=5.81 \times 10^{-3}$).


We plot the distribution of $h_{\rm in}/R_{\rm bar}$ for massive galaxies in Fig.~\ref{fig:mu0_hin2}(c) and that for less massive galaxies in Fig.~\ref{fig:mu0_hin2}(d). 
The distribution of $h_{\rm in}/R_{\rm bar}$ for galaxies with Bar/T$>$ 0.1 shows a narrow peak. This shows that longer bars reside in galaxies with a larger inner disc scale length among massive galaxies with Bar/T $>$ 0.1. 
Together with the results from Fig~\ref{fig:mu0_hin2}(a), this implies that longer bars reside in inner discs that show flattened radial surface brightness profiles. 
However, again, there is no clear relation among less massive galaxies or galaxies with Bar/T $\leq$ 0.1. 


Two dimensional galaxy decomposition fully takes into account the structural differences (e.g., ellipticity) of disc and bar in the galaxy model fit and the total mass of the various components is not kept constant in any way.
Therefore, our results do not come from just light (mass) reassignment between galaxy components in the fitting sense. 
For example, it is possible to have a model fit with a long, strong bar and a bright disc if there is such a galaxy.

\subsection{Quantifying the light deficit around the bar in inner discs}
\label{sec:res3}
\begin{figure*}
\centering
\includegraphics[width=16cm]{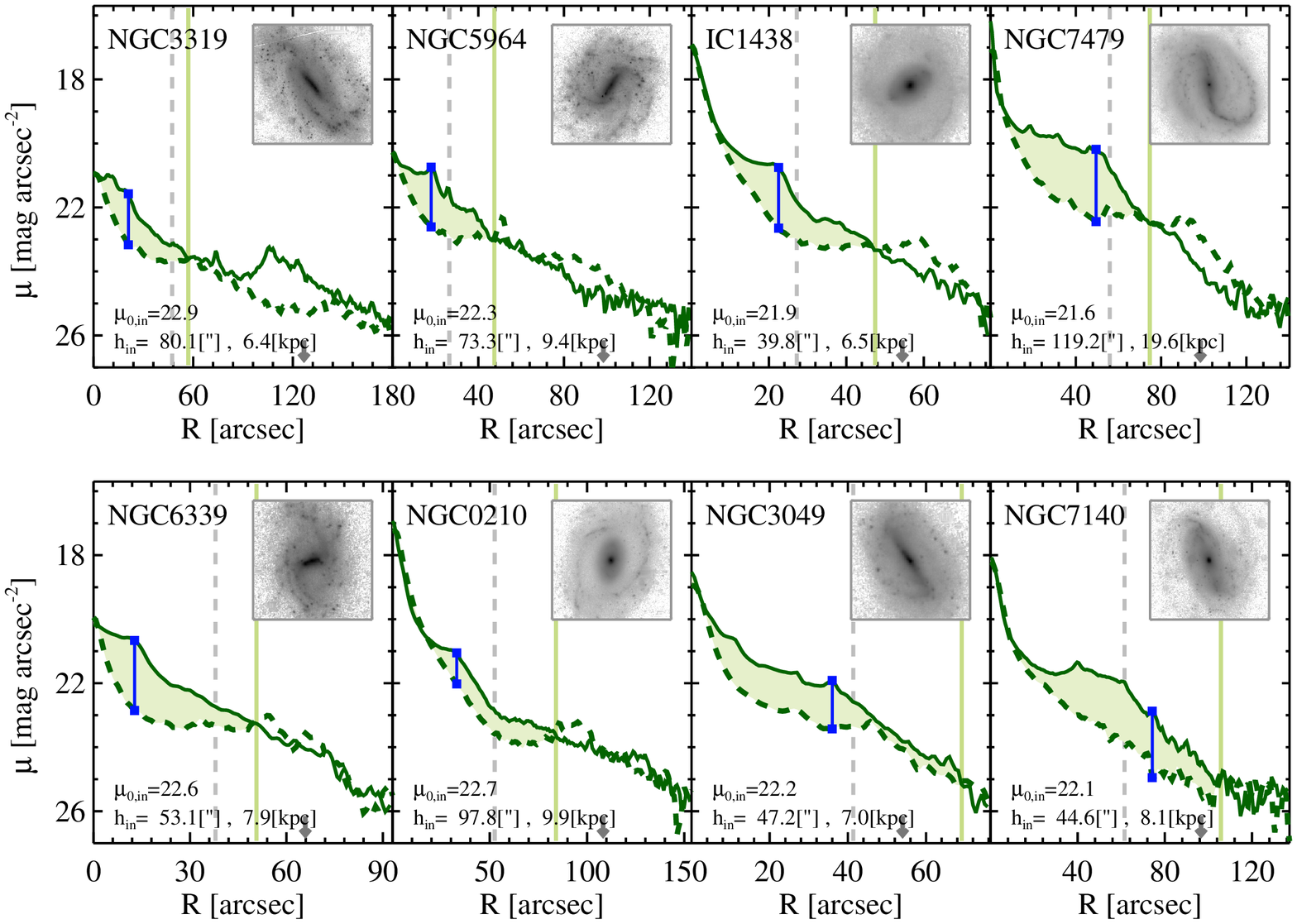}
\includegraphics[width=16cm]{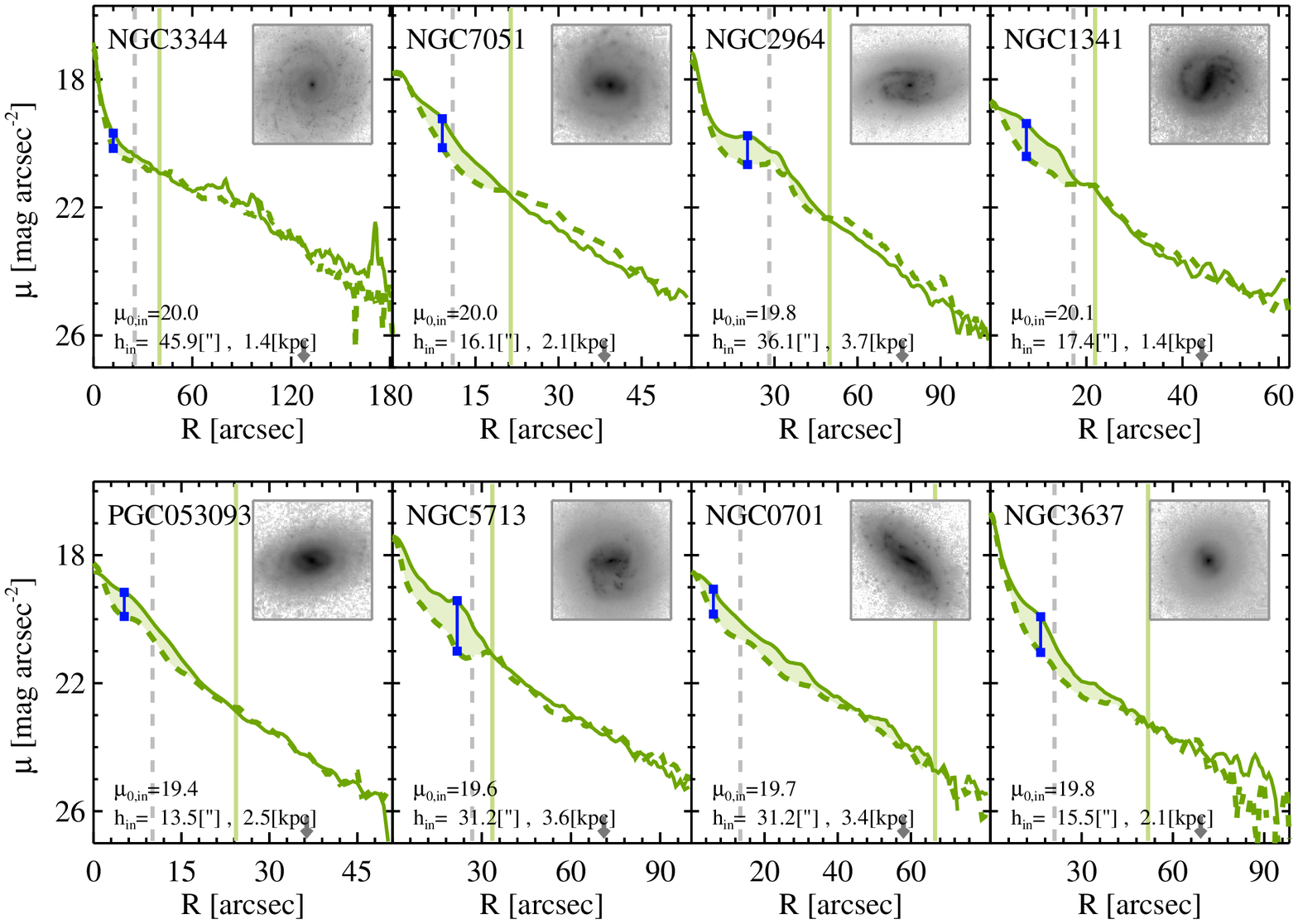}
\caption{Light profiles along the bar major axis (solid line) and minor axis (dashed line) taken from deprojected images. 
The inset image of each panel shows the \s4g image of the galaxy as observed, not deprojected. Each image covers the galaxy out to $r = 0.7 \times R_{\rm 25.5}$ that is denoted by the downward arrow.
The upper two panels represent galaxies with a large disc scale length and low inner disc central surface brightness, i.e. galaxies with flattened inner discs.
These galaxies are located in the lower-right part of Fig.~\ref{fig:mu0_hin}.
Galaxies in the lower two panels have a relatively smaller disc scale length and high inner disc central surface brightness, and thus lie in the upper-left part of Fig.~\ref{fig:mu0_hin}. Dashed grey vertical lines indicate the deprojected bar radius. Solid light green vertical lines indicate the radius where the two profiles cross ($R_{\rm cross}$). Vertical blue lines connecting two blue filled squares denote the measured Max($\Delta\mu$) (see text).
}
\label{fig:prof_mn_set8}
\end{figure*}

In the previous subsections, we have used the structural parameters ($\mu_{\rm 0, in}$ and $h_{\rm in}$) derived from our two-dimensional galaxy model fits. However, those are rather indirect quantities which need galaxy model fits to describe the deficit of light in the inner disc surrounding the bar. Here we take a different approach and devise an indicator for the light deficit in the inner disk using a non-parametric analysis to explore the relation between the bar and the light deficit in the inner disc.

Bars are non-axisymmetric features, and thus the light profiles of galaxies along the bar major and minor axes differ at $r<R_{\rm bar}$. In Fig.~\ref{fig:prof_mn_set8}, we plot light profiles of galaxies along the bar major and minor axes in solid and dashed lines, respectively. 
To obtain the bar major axis profile, first we measure the position angle of the bar for each galaxy, and rotate the image so that the bar is aligned horizontally. Then, we put a slit of 5.25 arcsec (7 pixels) width horizontally and calculate the mean surface brightness at each radius. 
The bar minor axis profile is calculated in the same way but by putting the slit vertically. 
Beyond the bar width, the bar minor axis profile basically traces the profile of the disc region.
Grey vertical dashed lines in Fig.~\ref{fig:prof_mn_set8} denote the bar length, and green vertical solid lines indicate the radii where the bar major and minor axes profiles meet or cross ($R_{\rm cross}$).
Galaxies in the upper two panels have a large $h_{\rm in}$ and a low $\mu_{\rm 0,in}$, i.e. they are located in the lower-right part of Fig.~\ref{fig:mu0_hin}. In turn, galaxies in the lower two panels have a smaller $h_{\rm in}$ and high $\mu_{\rm 0,in}$, and are located in the upper-left part of Fig.~\ref{fig:mu0_hin}. 

We define the Max($\Delta\mu$) [mag arcsec$^{-2}$] as the maximum difference between the surface brightness profiles along the bar major and minor axes. 
Therefore, by default, Max($\Delta\mu$) is positive for all barred galaxies and close to zero for weak, inconspicuous barred galaxies.
Max($\Delta\mu$) is a measure of the bar prominence in the sense that it contains the light from the bar above the disc.
Because bars become stronger by capturing disc stars, Max($\Delta\mu$) is also a measure of the light deficit in the inner disc, and Max($\Delta\mu$) also contains the light deficit below the average disc.
Therefore, we use Max($\Delta\mu$) as an indicator of the light deficit in the inner disc. We make use of images taken at 3.6 {\mum}, where dust extinction is minimal and the mass-to-light ratio variation due to the star formation history is not significant, thus the light deficit at 3.6 {\mum} can be directly translated to a mass deficit.

In Fig.~\ref{fig:prof_mn_set8}, vertical blue lines that connect two squares indicate the measurements of Max($\Delta\mu$). There is a tendency that Max($\Delta\mu$) is larger for galaxies in the upper two panels than for those in the lower two panels. This indicates that galaxies with large $h_{\rm in}$ and low $\mu_{\rm 0}$ (i.e. inner disc with flattened radial profile) tend to have large Max($\Delta\mu$). 

\begin{figure*}
\centering
\includegraphics[width=16cm]{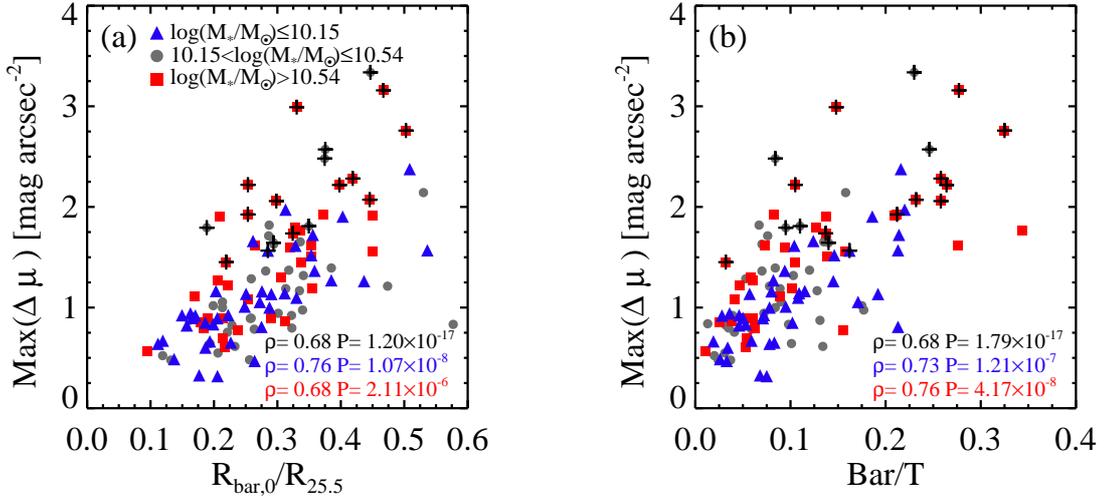} 
\caption{Relations between the maximum difference between surface brightness profiles along the bar major and minor axes, Max($\Delta \mu$), as shown in Fig.~\ref{fig:prof_mn_set8}, and selected bar properties. Max($\Delta \mu$) is plotted against 
(a): deprojected bar length normalized by $R_{\rm 25.5}$, and
(b): Bar/T.
Galaxies are colour-coded by their stellar mass. Galaxies are divided into three groups such that all groups have similar number of galaxies. Massive galaxies ($M_{*}/M_{\sun}>10.54$) are plotted in red squares, and less massive galaxies ($M_{*}/M_{\sun} \leq 10.15$) are plotted in blue triangles. Galaxies with $10.15<M_{*}/M_{\sun} \leq 10.54$ are plotted in grey circles. 
The Spearman correlation coefficient ($\rho$) and the statistical significance of the coefficient (P) are presented in each panel. Note that coefficients in black are for all the galaxies, those in red are for massive galaxies, and those in blue are for less massive galaxies.
Black crosses represent galaxies that lie above the dotted line in Fig 7(b). We visually  checked these galaxies and found that most of such galaxies have a bar that is either embedded in a lens/ovals or surrounded by an inner ring.
}
\label{fig:max_diff}
\end{figure*}

We now investigate how Max($\Delta\mu$) is related to bar parameters in more detail.
We plot Max($\Delta\mu$) as a function of normalized bar length ($R_{\rm bar}/R_{\rm 25.5}$) in Fig.~\ref{fig:max_diff}(a). 
Galaxies are colour-coded by their stellar mass. We divided galaxies into three groups: $\log(M_{\rm *}/M_{\sun}) \leq 10.15$ (40 galaxies), $10.15 < \log(M_{\rm *}/M_{\sun}) \leq 10.54$ (40 galaxies), and $\log(M_{\rm *}/M_{\sun})> 10.54$ (38 galaxies).  The boundaries of these groups are chosen so that they contain a similar number of galaxies.
The figure shows that longer bars tend to have an increased Max($\Delta\mu$). 
We plot Max($\Delta\mu$) against Bar/T in Fig.~\ref{fig:max_diff}(b), and they are also well correlated.
It is interesting to note that the majority of galaxies with Max($\Delta\mu$) $>2$ mag arcsec$^{-2}$ have Bar/T $>0.1$.
These two figures show that Max($\Delta\mu$) is a strong function of $R_{\rm bar}/R_{\rm 25.5}$ and Bar/T. This implies that conspicuous bars produce stronger light deficits in the inner part of the disc.

\begin{figure*}
\centering
\includegraphics[width=16cm]{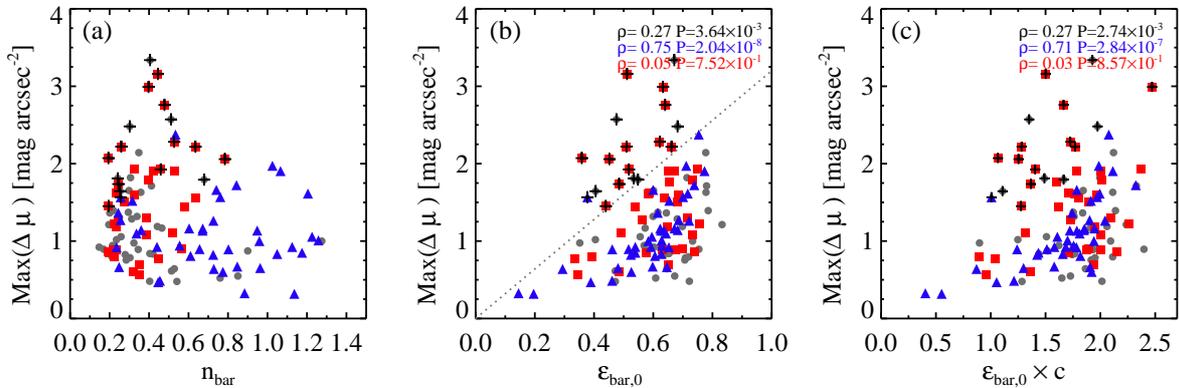} 
\caption{ Max($\Delta \mu$) versus 
(a): bar {\ser} index ($n_{\rm bar}$),
(b): deprojected bar ellipticity ($\varepsilon_{\rm bar,0}$), and 
(c): $\varepsilon_{\rm bar,0}$ multiplied by bar boxiness ($\varepsilon_{\rm bar,0} \times c$), both of which are a proxy for bar strength. Galaxies are colour-coded by their stellar mass as in Fig.~\ref{fig:max_diff}.
Black crosses are the galaxies that lie above the dotted line in Fig 7(b). These galaxies are also overlaid in crosses in Fig. \ref{fig:max_diff}. We visually  checked these galaxies and found that most of such galaxies have a bar that is either embedded in a lens/ovals or surrounded by an inner ring.
}
\label{fig:max_diff2}
\end{figure*}

Next, we show how bar {\ser} indices ($n_{\rm bar}$) are related to Max($\Delta\mu$) in Fig.~\ref{fig:max_diff2}(a). $n_{\rm bar}$ describes the shape of the bar radial surface brightness profile (for details, see \citealt{kim_15a}). In short, bars that have $n_{\rm bar} \sim$ 1 show exponential-like surface brightness profiles that resemble those of discs, while bars with $n_{\rm bar}\lesssim0.5$ show flat radial profiles. Because bars are formed from discs, it has been suggested that bars with $n_{\rm bar} \sim$ 1 are dynamically relatively young (i.e. have formed recently), while bars with $n_{\rm bar}\lesssim0.5$ are dynamically old (\citealt{kim_15a}).  
Fig.~\ref{fig:max_diff2}(a) shows that galaxies that have large Max($\Delta\mu$) -- i.e. larger than about 2 -- are predominantly massive galaxies with flat bars. However, apart from this, we do not find a clear connection between Max($\Delta\mu$) and $n_{\rm bar}$.

In Fig.~\ref{fig:max_diff2}(b) and (c), we plot the bar ellipticity ($\varepsilon_{\rm bar,0}$, deprojected) and the bar ellipticity multiplied by the bar boxiness ($\varepsilon_{\rm bar,0} \times c$). The bar boxiness, $c$, describes the face-on shape of bars (see \citealt{athanassoula_90} for a definition and \citealt{kim_15a} for more information on how to obtain its value).  We obtain $c$ treating bars as one component.
$\varepsilon_{\rm bar,0}$  has been shown to be a good measure of the bar strength for bars of roughly the same relative mass (\citealt{athanassoula_92a}), and has been extended to include all bars, independent of their mass, in observational studies as e.g. by \citet{gadotti_11}, who introduced also $\varepsilon_{\rm bar,0} \times c$ as a further measure of bar strength.
Fig.~\ref{fig:max_diff2}(b) indicates that except for massive galaxies, highly elongated bars tend to have larger Max($\Delta\mu$).
There are a number of points that lie well above the trend, and these are dominated by massive galaxies. These are the galaxies for which the light deficit in the disc surrounding the bar is very pronounced.
For galaxies located above the dotted line in Fig.~\ref{fig:max_diff2}(b) (overlaid with crosses), we have checked morphological classifications and analysis (\citealt{buta_15, herrera_endoqui_15}) and find that most of such galaxies have a bar that is either embedded in an inner lens (i.e., a lens that has the same major axis length as the bar) or surrounded by an inner ring. The exceptions are NGC3672 and NGC5964.
Our decompositions do not include models for the lens and the inner ring. Therefore, for these galaxies, our measures of bar ellipticity might be biased towards lower values. Removing these galaxies from the discussion or shifting them towards higher values of ellipticity strengthens the correlations between Max($\Delta\mu$) and $\varepsilon_{\rm bar,0}$ and $\varepsilon_{\rm bar,0} \times c$ in Figs.~\ref{fig:max_diff2}(b) and (c).
We also investigate whether removing those galaxies with crosses from Fig.~\ref{fig:max_diff}(a) and (b) makes any differences in the trend by checking Spearman correlation coefficients. We find that even after omitting out those galaxies, we still find clear correlations.

If a galaxy has a prominent inner lens around the bar that fills the disc region within the bar radius, such inner lens might hinder us from measuring the light deficit of the inner disc. For such galaxies, Max($\Delta\mu$) might be small because even along the bar minor axis, inner lens brings up the surface brightness profile (e.g., NGC0210, NGC3637 in Fig \ref{fig:prof_mn_set8}.). In our sample, only $\sim$11\% of galaxies have either bar lens and/or ring lens (\citealt{buta_15}), thus the results we obtain in Section \ref{sec:res3} remain intact.

In addition to Max($\Delta\mu$), another proxy for the light deficit in the inner disc would be the area between the galaxy light profiles along the bar major and minor axes (see green shaded area in Fig.~\ref{fig:prof_mn_set8}). 
We also estimated this area inside of $R_{\rm cross}$ to check how it correlates with the other parameters presented in Fig.~\ref{fig:max_diff} and Fig.~\ref{fig:max_diff2}. We find that the correlations are similar or less strong compared to the corresponding correlations with Max($\Delta\mu$). This is because any errors in producing deprojected images (e.g., ellipticity and position angle) can affect these areas considerably, whereas Max($\Delta\mu$) is relatively less affected by such errors.

In summary, we find that Max($\Delta\mu$) is strongly related to bar length, Bar/T, and also to the bar ellipticity (but to a lesser degree). This demonstrates that the light deficit in the inner disc is connected to the bar.


\section{Evolution of the light profile in simulated barred galaxies}
\begin{figure*}
\centering
\vspace{0.5cm}
\includegraphics[width=7cm]{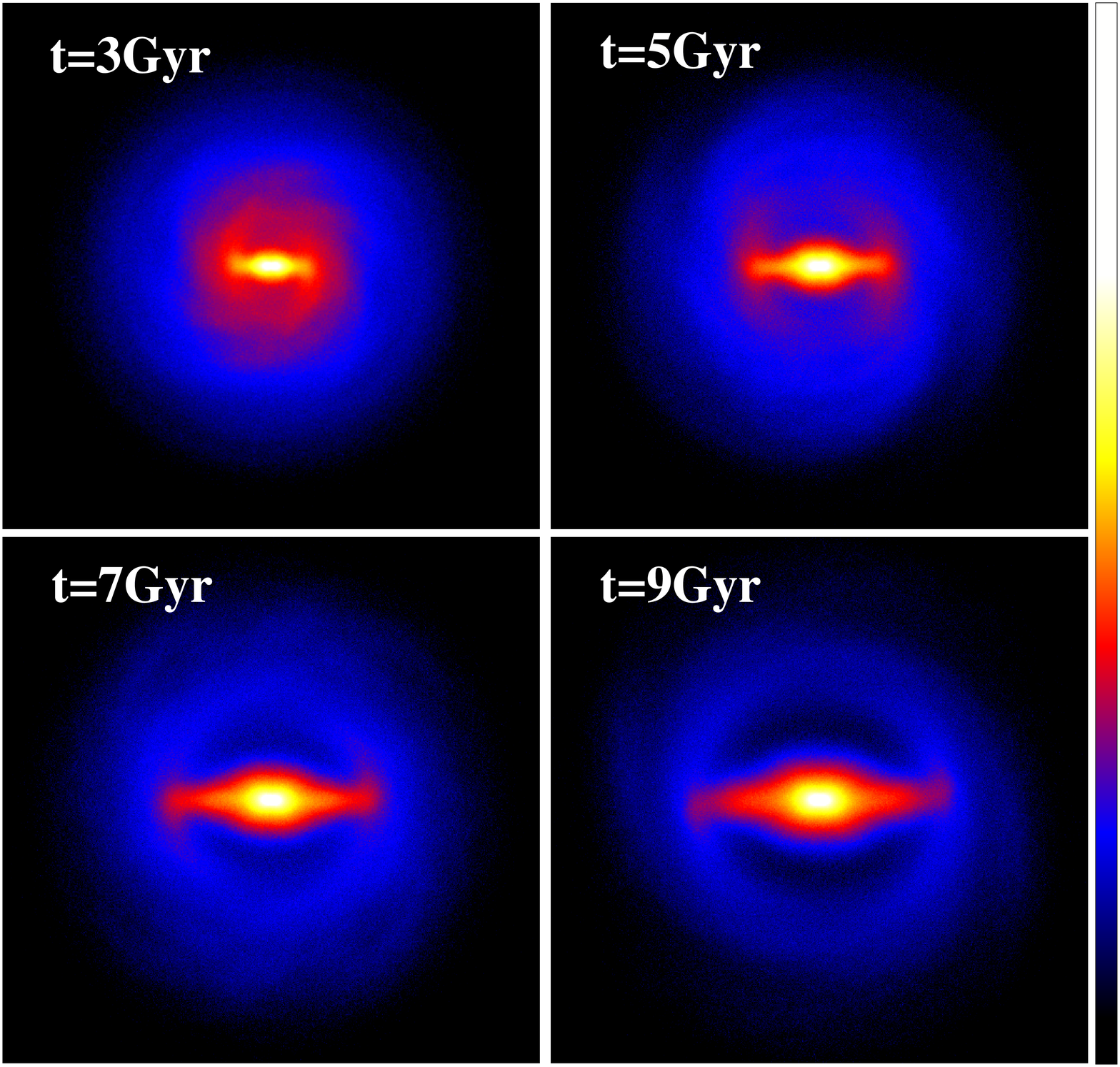}
\hspace{1.0cm}
\includegraphics[width=8cm]{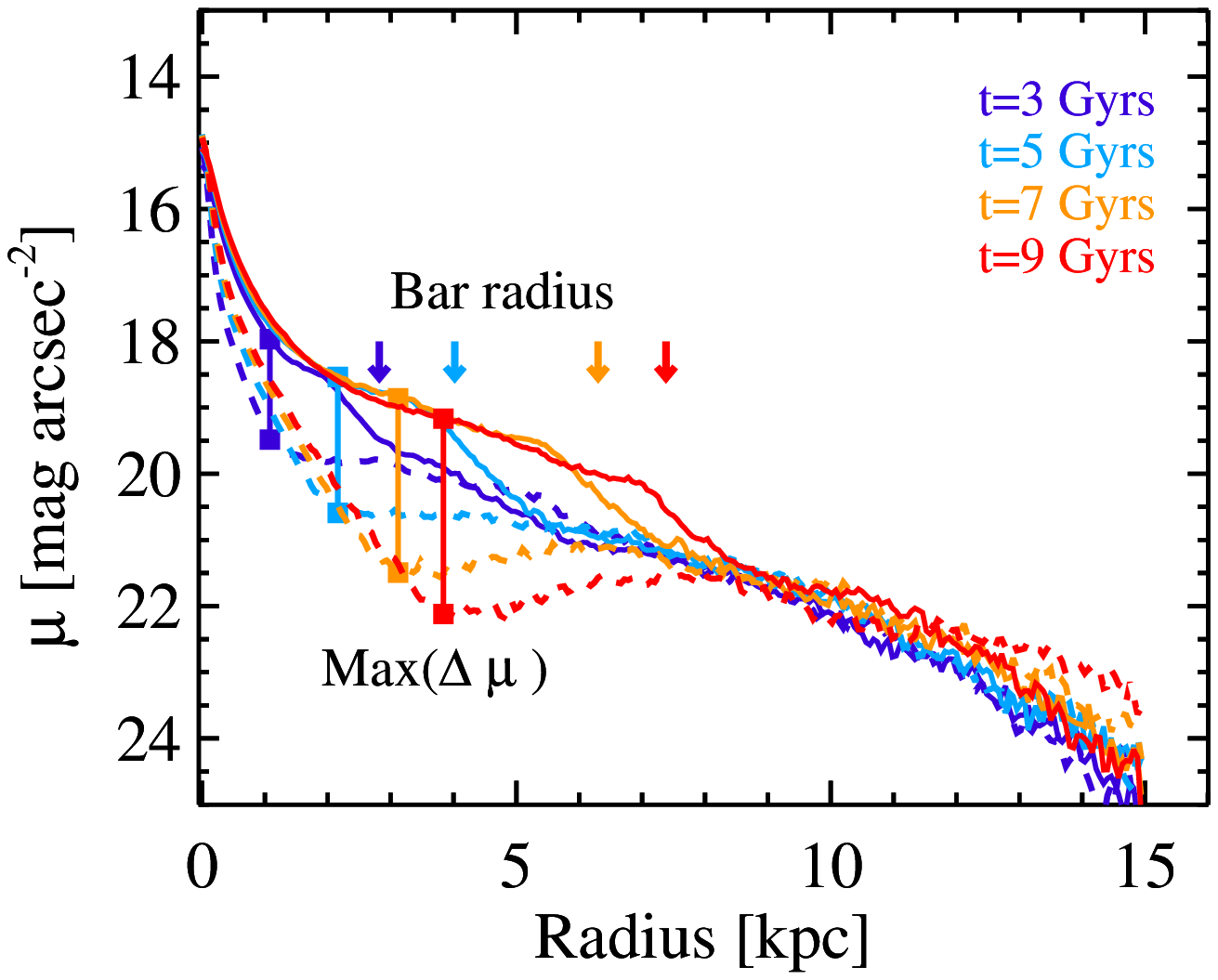} 
\caption{Left panel: Snapshot images of the simulated galaxy from \citet{athanassoula_13} (gtr116) at $t=3$, 5, 7 and 9 Gyr. All images span 30 $\times$ 30 kpc, and are displayed to have the same stretch. The light deficit is more pronounced at $t=7$ and 9 Gyr, as compared to $t=3$ and 5 Gyr.  Right panel: Light profiles of the simulated galaxy at the four different times. Solid lines represent profiles along the bar major axis while dashed lines represent profiles along the bar minor axis.  Bar radii are indicated with downward arrows, and Max($\Delta \mu$) for each time is plotted with a vertical line connecting two filled squares.}
\label{fig:gtr116_img}
\end{figure*}

With observational data, we cannot directly trace the evolution of a given galaxy, but only infer their evolution globally, by studying galaxies with e.g. various Hubble types and masses, and/or applying indirect or statistical methods.
Thus, simulations come to play an important role as they easily provide us information on how individual simulated galaxies evolve.
We now make use of simulations to trace how the mass profiles of the simulated galaxy change with time.
We use snapshot images from an $N$-body simulation of a disc galaxy by \cite{athanassoula_13}.
These include gas and its physics, namely star formation feedback and cooling. They are also fully self-consistent, including a live, responsive halo which allows the angular momentum exchange within the galaxy to be correctly modeled (see \citealt{athanassoula_02b, athanassoula_13} for a comparison of the effects of a rigid and a live halo on bar growth and evolution). 
We show the image obtained from all stellar particles, independent of their age, of simulation ``gtr116'' at $t=3$, 5, 7 and 9 Gyr in Fig.~\ref{fig:gtr116_img}. 
All images have the same physical scale (30 $\times$ 30 kpc) and are displayed to have the same stretch (scale and contrast). 
From $t=3$ to 9 Gyr, the light deficit in the inner disc becomes more prominent and the bar becomes longer. We plot the bar major and minor axis profiles in the right panel of Fig.~\ref{fig:gtr116_img} for the simulated galaxy at $t=3$, 5, 7 and 9 Gyrs, just like we did in Fig.~\ref{fig:prof_mn_set8} for the real galaxies in our sample. Profiles along the bar major and minor axes are plotted as solid and dashed lines, respectively. 
The figure shows that as the galaxy evolves, the minimum of the dip in the bar minor axis profile reaches fainter density (luminosity) levels, so that Max($\Delta\mu$) increases by $\sim$ 1.4 mag arcsec$^{-2}$ from $t=3$ to 9 Gyr, and the light deficit in the inner disc becomes more prominent. In addition to the increase of Max($\Delta\mu$), the bar becomes also longer with time.
These results confirm previous results by AM02, but extend them and supersede them in that they include gas (and its physics, namely star formation, feedback and cooling),  which is a central protagonist in secular evolution.

\section{Discussion}
\label{sec:discussion}

We show above that, when considering massive galaxies ($M_{*}/$\Msun$> 10^{10}$), longer bars tend to reside in inner discs that show flattened surface brightness profiles, when Bar/T is above 0.1 (see Section~\ref{sec:res_surfb_hin} and ~\ref{sec:res2}). This tendency is relatively weak for less massive galaxies. However, devising a more direct measure of the light deficit in the inner disc, Max($\Delta \mu$), we find that such deficit clearly shows up also among less massive galaxies (Sect.~\ref{sec:res3}). The relation between Max($\Delta \mu$) and bar length and Bar/T clearly holds throughout the mass range explored in this study ($10^9 - 10^{11}$ {\Msun}).
What is the origin of these light deficits?

Numerical simulations find that as a barred galaxy evolves, the bar loses angular momentum to the outer disc or halo (see \citealt{athanassoula_13_book} for a review and references therein), and the bar becomes more elongated, changing the shape of its main orbital families and making them thinner and more extended by trapping disc stars near the bar in orbits belonging to the bar (\citealt{athanassoula_03}).
As a consequence, we conjecture that those captured disc stars lead to the light deficit in the inner disc, and the inner disc thus shows a flattened radial surface brightness profile (fainter $\mu_{\rm 0}$ and larger $h_{\rm in}$). 
In some cases, the light deficit becomes considerably pronounced and the inner disc becomes very faint within the bar radius, as can be seen in Figs.~\ref{fig:deficit_light} and \ref{fig:gtr116_img} \citep[see also][]{gadotti_03}. 

Our quantification of the light deficit, given by the parameter Max($\Delta\mu$), proved to be very elucidative. Max($\Delta\mu$) might be small, or, more probably, absent when there is no bar. But once the bar forms, Max($\Delta\mu$) should increase and become significant. 
By capturing more and more disc stars into the bar, the bar becomes longer and at the same time, the light deficit in the inner disc becomes more pronounced, and thus Max($\Delta\mu$) increases.

As we find that the galaxies with a longer bar and higher Bar/T have larger Max($\Delta\mu$), our results are consistent with theoretical expectations, strongly suggesting that indeed the light deficit observed in the inner discs is produced by bars. In addition, we also find that the vast majority of galaxies with Max($\Delta\mu$) above 2 are in our highest bin of stellar mass, and have bars with flat radial profiles. This is consistent with our picture that bars form first in more massive galaxies, and thus more massive galaxies today have more evolved bars \citep{sheth_12}. And it is also consistent with the idea that massive galaxies mostly host evolved bars that show flatter radial profiles \citep{kim_15a}. Thus, our results provide direct observational evidence for such bar-driven secular disc evolution.


We have shown that the disc material is captured onto bar orbits as the galaxy evolves. Another important result to emphasize is that the material is taken out of specific locations of the disc, and these specific locations agree with the results from simulated galaxies that have undergone bar driven secular evolution. We will discuss these points, the corresponding dynamical implications and their links to observations in a forthcoming paper based on simulations. 


There are some caveats to note. In this study, bar radii and bar ellipticities are deprojected analytically assuming that the bar shape can be approximated by an ellipse. Although bars can be approximated by ellipses, a more elaborate description with generalized ellipses (\citealt{athanassoula_90}, also the appendix in \citealt{athanassoula_14}) shows that bars are slightly more boxy. Thus, any differences between the shape of the bar and the approximated ellipse might produce uncertainty in the deprojection. Also the uncertainties in the inclination angle of the galaxy and position angle of the line of nodes will produce uncertainties in the deprojection.

In our decompositions we used one single ellipsoidal component to describe the bar while bars have a complex shape. The outer part of the bar is both horizontally and vertically thin and an inner part of the bar is thick in both directions (e.g., \citealt{athanassoula_05a}). 
When seen face-on, the inner component is called barlens, a component first recognised and classified in \citet{laurikainen_11}, but seen edge-on are known as a boxy/peanut/X (\citealt{laurikainen_14, athanassoula_15}).
To take into account the bar geometry in more detail, we need more sophisticated decompositions that include two independent components of bars (\citealt{laurikainen_05, laurikainen_14, athanassoula_15}. The latter paper discusses further shortcomings of assuming a bar as a single ellipsoidal). 
If bars had been modelled with two components, we might have been able to put better constraints on whether there is a relation between Max($\Delta\mu$) and the profile of each component of the bar.
However, such decompositions are beyond the scope of this paper, and will be considered in the forthcoming theoretical paper.

\section{Summary and Conclusions}
Bars act as a driving force for the evolution of their host galaxies. 
With the aim of assessing the impact of bar-driven secular evolution on discs,
we used 3.6 {\mum} images of 118 nearby barred galaxies from the {\s4g} with type II (down-bending) radial surface brightness profiles. We investigated how the properties of bars are related to those of the inner parts of their host discs. In particular, we investigated the origin of the light deficits often observed in the part of inner discs surrounding the bar, within the bar radius \citep[see e.g.][for earlier discussions]{gadotti_08}. Our main results can be summarized as follows:

\begin{itemize}

\item Among massive galaxies with a prominent bar (Bar/T$>0.1$), there is a clear trend that longer bars reside in more flattened inner discs (larger inner disc scale length and lower central surface brightness) than shorter bars do.
Such galaxies often show the light deficit around the bar in the inner part of the disc.

\item To better understand the relation between the bar and the light deficit in inner discs, we quantify the light deficit. We measure the maximum difference between the surface brightness profiles along the bar major and minor axes, Max($\Delta \mu$). 
As it measures the light above the disc, Max($\Delta \mu$) is a measure of the bar prominence. Because bars evolve by capturing disc stars, Max($\Delta \mu$) is also a indicator for the light deficit in the inner disc. 
We find that Max($\Delta \mu$) is strongly related to the bar size and to Bar/T, so that the light deficit is directly proportional to bar size and to how conspicuous the bar is.

\item By studying a time sequence of snapshots from the evolution of a simulated barred galaxy, we find that as the bar evolves, it becomes longer and the light deficit in the inner disc becomes more pronounced. This can be understood by the fact that as a barred galaxy evolves, the bar loses angular momentum and becomes longer and more massive by trapping nearby disc stars onto bar orbits. Therefore, the light deficit is produced as a consequence of the capture of disc stars by the bar, which are thus removed from the inner part of discs.

\end{itemize}

The observed correlations between the light deficit and bar size and Bar/T (Fig.~\ref{fig:max_diff}) are consistent with the picture drawn from the analysis of the evolution of a simulated barred galaxy (Fig.~\ref{fig:gtr116_img}), in that bars grow longer and more conspicuous by capturing nearby disc stars. Based on these results, we therefore propose that the light deficit often observed in the part of the inner discs within the bar radius is produced by bars. This is direct evidence for bar-driven secular evolution in galactic discs, and a strong indication that bars are actively involved in shaping the mass distribution of their host galaxies.

\section*{Acknowledgements}
{\it Facilities:} {The {\em{Spitzer}} Space Telescope}
\\
\\
The authors thank the \s4g team for their effort in this project.
T.K., K.S., acknowledge support from the National Radio Astronomy Observatory, which is a facility of the National Science Foundation operated under cooperative agreement by Associated Universities, Inc.  We are grateful for the support from NASA JPL/Spitzer grant RSA 1374189 provided for the {\s4g} project.
T.K. and M.G.L. were supported by the National Research Foundation of Korea (NRF) grant funded by the Korea Government (MEST) (No. 2012R1A4A1028713).
D.A.G., E.A., and A.B. thanks the Marie Curie Actions of the European Commission (FP7-COFUND) for the funding.
E.A. and A.B. thank the Centre National d'Etudes Spatiales (CNES) and the
``Programme National de Cosmologie and Galaxies'' (PNCG) of
CNRS/INSU, France for financial support. They also acknowledge financial support from the People Programme (Marie Curie Actions) of the European Union's FP7/2007-2013/ to the DAGAL network under REA grant agreement No. PITN-GA- 2011-289313. E.A. acknowledges the use of HPC resources
from GENCI-TGCC/CINES (Grants 2013 - x2013047098 and 2014 - x2014047098).

This research is based on observations and archival data made with the {\em{Spitzer}} Space Telescope, and made use of  the NASA/IPAC Extragalactic Database (NED) which are operated by the Jet Propulsion Laboratory, California Institute of Technology under a contract with National Aeronautics and Space Administration (NASA). We acknowledge the usage of the HyperLeda database (http://leda.univ-lyon1.fr).

\bibliographystyle{mnras}
\bibliography{tkim_bar}
\clearpage


\bsp    
\label{lastpage}
\end{document}